\documentstyle[aps,epsf,rotating]{revtex}

\begin{document}

\preprint{}

\title{Evidence for Scale-Scale Correlations in the 
                    Cosmic Microwave Background Radiation}
\author{Jes\'{u}s Pando$^1$, David Valls--Gabaud$^{1,2}$ and Li-Zhi Fang$^3$}
\address{$^1$ UMR 7550 CNRS, Observatoire de Strasbourg,
         11 Rue de l'Universit\'e, 67000 Strasbourg, France}
\address{$^2$ Institute of Astronomy, Madingley Road, Cambridge CB3 0HA, UK}
\address{$^3$ Department of Physics, University of Arizona, 
         Tucson, AZ 85721, USA}
\date{\today}
\maketitle

\begin{abstract}
We perform a discrete wavelet analysis of the COBE-DMR 4yr sky maps and 
find a significant scale-scale correlation on angular scales from about
11 to 22 degrees, only in the DMR face centered on the North Galactic Pole. 
This non-Gaussian signature does not arise either from the known foregrounds 
or the correlated noise maps, nor is it consistent with upper limits on the 
residual systematic errors in the DMR maps.
Either the scale-scale correlations are caused by an unknown foreground 
contaminate or systematic errors on angular scales as large as 22 degrees, 
or the standard  inflation plus cold dark matter paradigm is ruled out at 
the $> 99\%$ confidence level.
\end{abstract}

Most attempts at quantifying the non-Gaussianity in the cosmic microwave
background radiation are motivated by the belief that non-Gaussianity can
distinguish inflationary models of structure formation from topological
models. While standard inflation predicts a Gaussian distribution of
anisotropies \cite{infl}, spontaneous symmetry breaking produces topological
defects whose networks create non-Gaussian patterns on the microwave
background radiation on small scales\cite{def}. Minute non-Gaussian features 
can however be generated by gravitational waves \cite{gravw} or by the 
Rees-Sciama \cite{rs} and Sunyaev-Zeldovich effects. 

It is generally held that cosmic gravitational clustering can be roughly 
described by three r\'egimes: linear, quasi-linear, and fully developed
nonlinear clustering. Whilst quasi-linear and non-linear clustering
induce non-Gaussian distribution functions, if the initial density 
perturbations are Gaussian, scale-scale correlations and other non-Gaussian 
features of the density field can not be generated during the linear 
r\'egime. Hence the linear r\'egime is best suited to study the primordial 
non-Gaussian fluctuations. Since the amplitudes of the cosmic temperature 
fluctuations revealed by COBE are as small as $\Delta T/T\simeq 10^{-5}$,
the gravitational clustering should remain in the linear r\'egime on scales 
larger than about 30 $h^{-1}$ Mpc and at redshifts higher than 2.
Current limits on non-Gaussianity from galaxy surveys probe redshifts smaller 
than about 1 \cite{sp96}. Interestingly, at redshifts between 2 and 3, and
scales on the order of 40 to 80 $h^{-1}$ Mpc, there are positive detections
of  scale-scale correlations
in the distribution of Ly$\alpha$ 
absorption lines in quasar spectra \cite{p98a}. These clouds are likely to be 
pre-collapsed and continuously distributed intergalactic gas clouds,
and are therefore fair tracers of the cosmic density field, especially on 
large scales \cite{bi97}. 
This may indicate that the primordial fluctuations are scale-scale correlated.

While on small angular scales ($\ell \simeq 150$) there may be some
indications of non-Gaussianity \cite{gfe98}, studies by traditional 
non-Gaussian detectors have concluded that there is no evidence of 
non-Gaussianity in the cosmic temperature fluctuations on large scales 
\cite{ko96}. (See however \cite{fmg98}.) This does not rule out the 
existence of scale-scale correlations. Because each non-Gaussian feature 
is non-Gaussian in its own way, there is no single statistical indicator for 
the existence of non-Gaussianity in data. For instance, there are models of 
scale-scale coupling which lead to a density field with a Poisson distribution
in its one-point distribution function, but that are highly scale-scale 
correlated \cite{g95a}. In this case, all statistics based on the one-point 
functions will fail to detect the scale-scale correlations, that is, they 
will miss the non-Gaussianity. As yet, the scale-scale correlations of the 
cosmic temperature fluctuations have not been searched for in any available 
data set. 
It is the intent of this Letter to
probe for the scale-scale correlations in the COBE-DMR 4-year sky maps, and, 
as an example, show that this measure is effective in testing models of the
initial density perturbations. 
In contrast with other techniques, such as
the bispectrum, \cite{bispec}, higher order cumulants \cite{cumul}, 
Minkowski functionals \cite{wiko97}, or double Fourier analysis \cite{lewin98},
scale-scale correlations are localized, and can localize the areas on
the sky where the signal comes from, and with a resolution that depends
on the scale considered.

The scale-scale correlations are conveniently described by the discrete
wavelet  transform (DWT) \cite{p98a,f97a}.
Considering a 2-dimensional temperature (or density) field $T({\bf x})$, 
where ${\bf x}=(x_1,x_2)$, such that
$0 \leq x_1,x_2 \leq L$, the DWT scale-space decomposition of the contrast 
$\Delta T({\bf x})/T$ is  
\begin{equation}
\frac{\Delta T}{T} = \sum_{j_1=0}^{J_1-1}\sum_{j_2=0}^{J_2-1} 
\sum_{l_1=0}^{2^{j_1}-1}\sum_{l_2=0}^{2^{j_2}-1} 
\tilde{\epsilon}_{j_1,j_2;l_1,l_2} \psi_{j_1,j_2; l_1, l_2}({\bf x}), 
\end{equation}
where  $\psi_{j_1,j_2; l_1, l_2}({\bf x})$ ($j_1,j_2 = 0, 1, 2..$ and
$l_1=0, 1...2^{j_1}-1$, $l_2=0, 1...2^{j_1}-1$) are the complete and orthogonal
wavelet basis \cite{dau92}. 
The indexes $(j_1,j_2)$ and $(l_1,l_2)$
denote the scale $(L/2^{j_1},L/2^{j_2})$ and position
$(Ll_1/2^{j_1},Ll_2/2^{j_2})$ in phase space and $J_1$ and $J_2$ are the 
smallest scales possible (i.e., one pixel). The wavelet 
basis function, 
$\psi_{j_1,j_2; l_1, l_2}({\bf x})$, is localized at the phase space point
$(j_1,j_2;l_1,l_2)$ and the wavelet coefficients 
$\tilde{\epsilon}_{j_1,j_2;l_1,l_2}$ measure the 2-D perturbations at the 
phase space point $(j_1,j_2;l_1,l_2)$. 
To be specific, we will use the Daubechies 4 wavelet in this paper, although
the results are not affected by this choice so long as a compactly supported
wavelet basis is used.

To measure correlations between scales $(j,j)$ and $(j+1, j+1)$, we
define
\begin{equation}
C^{p,p}_j = \frac{2^{2(j+1)}\sum_{\bf{l}=0}^{2^{j+1}-1}
\tilde{\epsilon}^p_{j;{\bf [l/2]}} \; \tilde{\epsilon}^p_{j+1;{\bf l}}}
{\sum \tilde{\epsilon}^p_{j,{\bf [l/2]}} \sum \tilde{\epsilon}^p_{j+1;{\bf l}}}
\end{equation}
where $p$ is an even integer, $ {\bf l} \equiv (l_1,l_2)$, and the 
[\hspace{2mm}]'s denote the integer part
of the quantity. Because $Ll/2^j = L2l/2^{j+1}$, the position 
$l$ at scale $j$ is the same as the positions $2l$ and $2l+1$ at scale 
$j+1$. Therefore, 
$C_j^{p,p}$ measures the correlation between scales 
at the {\it same} physical point. 
For Gaussian fields, $C_j^{p,p}=1$. $C_j^{p,p}>1$ corresponds to a positive
scale-scale correlation, and $C_j^{p,p}<1$ to the negative case. One can also 
show that a $C_j^{p,p} >1$ field cannot be produced by a $C_j^{p,p}<1$ 
distribution in a Gaussian background.   

It is also possible to define the more ``standard" non-Gaussian measures with
the wavelet coefficients.  Namely we define the third and fourth order
cumulants as
\begin{equation}
S_j \equiv \frac{1}{(M_j^2)^{3/2}}M^3_j, \quad \quad 
K_j  \equiv   \frac{1}{(M_j^2)^2} M^4_j -3,
\quad \quad \quad \quad{\rm where} \quad \quad
M^n_j \equiv \frac{1}{2^{2j}} \sum_{l_1,l_2=0}^{2^j-1}
(\tilde{\epsilon}_{j,j:l_1, l_2} -
\overline{\tilde{\epsilon}_{j,j;l_1,l_2}})^n,
\end{equation}
and $\overline{\tilde{\epsilon}_{j,j;l_1,l_2}}$ is the ensemble average 
(simulated samples) or the average over $(l_1, l_2)$ (real data).

The COBE-DMR data is formatted such that the entire sky is projected onto a
cube with each of its 6 faces  pixelized into $2^{10}$ approximately 
equal-area pixels. Although one could think of performing a spherical wavelet 
analysis directly on the sky, the current format is ideal for a direct 2-D DWT 
analysis. The  pixels of each face can be labeled by $(j_1,j_2)$ with 
$0 \leq j_1,j_2 \leq 5$, and $(l_1, l_2)$ with $0 \leq l_1 \leq 2^{j_1}$ and 
$0 \leq l_2 \leq 2^{j_2}$.  The scale $j$ corresponds to angular scale 
$2.8 \times 2^{5-j}$ degrees. In this way, one can
analyze each face individually. This is important, as we can reduce the
influence of the galactic foreground contamination by selecting the faces in the
direction of galactic poles. The galactic plane stretches across Faces 1 - 4
(in galactic coordinates) of the projected cube, while Faces 0 and 5 are
relatively free of galactic interference. We will concentrate on these two
faces since the standard galactic cut at $|b|$ = 20 degrees implies that the 
other faces will be significantly contaminated.

Before attempting to measure the non-Gaussianity in the DMR maps, we should 
test for possible contamination due to various kinds of noise. 
A typical example of non-Gaussianity caused by noise is Poissonian noise. 
Fortunately, this type of non-Gaussianity can be properly handled by the higher 
order DWT cumulant spectra  \cite{f97a}.
To quantify any non-Gaussianity due to DMR noise, we generate
1000 realizations of the temperature maps for a typical CDM model with
parameters $\Omega_0=1$, $h = 0.5$, and $\Omega_b = 0.05$ and generate
the appropriate sky maps at the DMR resolution \cite{cmap}. 
To these maps, we linearly add noise to each pixel by drawing from a Gaussian
distribution with the pixel dependent variance given by the two
different foreground removal techniques, DCMB and DSMB (see \cite{ben92} for 
details.) 

Previous non-Gaussian studies using a genus method and other
statistics, have found the four year DMR data to be consistent with a Gaussian 
field \cite{ko96}. The evaluation of the genus at different smoothing angles 
is similar to the DWT scale decomposition which is also based on smoothing 
on various angular scales and suggests that the DWT cumulant spectra
should give similar results. 
The results for $S_j$ and $K_j$ of the COBE-DMR foreground removed maps
and the CDM model are shown in Fig. 1.  
Using the 1000 realizations of 
the CDM model, we construct the probability distribution for both $S_j$ and
$K_j$. Fig. 1 gives the most probable values of $S_j$ and $K_j$ for the
CDM model with the error bars corresponding to the 95\% probability of drawing 
$S_j$, $K_j$ from the CDM model. Fig. 1 also shows that $S_j$ and $K_j$ for
the DCMB and DSMB data are safely within the 95\% range. Therefore, one can 
conclude that no significant non-Gaussianity can be identified from the third 
and fourth order cumulants. This result is consistent with the genus results.
Note that contrary to previous studies,  we can study the six faces of the
cube separately. 
Fig. 1  shows that both $S_j$ and 
$K_j$ are isotropic with respect to the face 0 and 5.
 
We can now proceed to the scale-scale correlations. We list the most probable
values of $C_j^{2,2}$ for the CDM+DSMB maps in Table 1. Similar results are 
obtained for the CDM+DCMB maps.
Because the CDM model is a Gaussian model, all $C_j^{2,2}$ are about equal 
to 1 as expected. At any scale $j$, $C_j^{2,2}$ is about the same for face 0 
and 5.
Therefore, the noise from the two foreground removed DMR maps does
not cause significant spurious scale-scale correlations and are
thus suitable for a scale-scale correlation analysis. 
At the very least the sample is good for a comparison between
observed scale-scale correlations with the CDM model.

The results for $C_j^{(2,2)}$ for the COBE-DMR foreground removed maps are 
plotted in Fig. 2 and tabulated in Table 1. 
The behavior of $C_1^{(2,2)}$ is markedly 
different from $S_j$, $K_j$, or the CDM+DSMB results. First, $C_1^{(2,2)}$ for 
face 0 cannot be 
drawn from the CDM model with a probability greater than 99\%. Second, 
$C_1^{(2,2)}$ is not isotropic, showing a difference between faces 0 and 5.
The DCMB maps show the same behavior.

$C_1^{(2,2)}$ describes the correlation between perturbations on angular
scales of $\simeq 22$ and $\simeq 11$ degrees, which corresponds to comoving
scales larger than about 100 $h^{-1}$ Mpc. Because the wavelets are orthogonal, 
$C_1^{(2,2)}$ cannot be changed by adding any abnormal process on angular 
scales less than $\simeq$ 10 degrees. The ``surprisingly" large value for
$C_1^{(2,2)}$ cannot be explained by any non-Gaussian process on small scales.
We have shown that the errors of  the foreground removed DMR maps cannot 
contribute to $C_1^{(2,2)}$. 

We also checked for possible contributions to 
the non-Gaussianity from systematics by doing a similar analysis on the 
systematic error maps.
It is unlikely that the detected non-Gaussianity comes from the 
systematics since the non-Gaussianity is on the order of
$\simeq 10^{-5}$ K, while the contribution to the anisotropy from the
systematics  is estimated to be on order $\simeq 10^{-6}$ K \cite{ko96b}.
The analysis of the combined systematic error maps 
confirms that $C_j^{2,2}$ is solidly in the Gaussian r\'egime, i.e., 
$C_1^{2,2} = 1.247 \pm 0.375$.  Moreover, these angular
scales are larger than the resolution of the DMR instrument. Therefore,
unless there are very local foreground contaminations which are overlooked 
by the two foreground removal methods, the high value for and the
anisotropy in $C_1^{(2,2)}$ is cosmological. 

To check if there could be large-scale foreground correlations overlooked
by the COBE-DMR subtraction technique, we performed the same analysis on
the dust maps generated by a careful combination of IRAS and DIRBE 
data \cite{sfd98}. Depending on the method used to obtain an averaged
value for the color excess E(B-V) on a DMR pixel, the $C_1^{2,2}$ values
range from $0.572 \pm 0.747$ to $0.630 \pm 0.617$. Although these
maps show
small-scale structure, when averaged over scales larger than the 
DMR pixels (2.8 degrees) any non-Gaussian fluctuation disappears.
In addition we checked the possibility that the non-Gaussianity
was due to anisotropies in the synchrotron emission by analysing an all-sky
map at 408 MHz \cite{haslam}.
A visual inspection of this map shows a structure extending from the galactic 
plane on to the North Galactic Pole.  
However, using a map projected in the same way as the DMR maps we
obtain $C_1^{2,2} = 0.069$, which is much less than the value
$C_1^{2,2} = 1.008 \pm 0.342$ obtained by 1000 bootstrap random realizations.
As mentioned above, $C_j^{2,2}>1$ cannot come from a superposition of a 
distribution with $C_j^{2,2}<1$ in a Gaussian background. 
Thus the scale-scale correlation detected in the COBE-DMR data is not a result 
of this signal.  Additionally, none of the individual frequency maps nor a 
linear combination consisting of the 53 GHz and 90 GHz frequencies maps 
show $C_1^{2,2} > 1.5$. Since these maps do not contain the foreground
subtractions, this result implies that if the cause of the signal in face 0
is foreground, it is incoherent.

As a final check, we also looked at the correlated noise maps in COBE-DMR 
\cite{Lin:1994}.  The individual correlated noise maps of the frequencies were 
checked for scale-scale 
correlations and once again, $C_1^{2,2}$ was solidly in the Gaussian r\'egime 
with  $C_1^{2,2} = 1.007 \pm 0.440$ for the 31 GHz channel, 
$C_1^{2,2} = 0.896 \pm 0.308$ for the 53 GHz channel, and 
$C_1^{2,2} = 0.935 \pm 0.310$ at 90 GHz.  

If indeed we have eliminated all non-cosmological sources that could account
for this signature and if the signal is not just a statistical fluke (since 
there is still a 1\% chance of this occuring), then the only conclusion 
left is that the correlation
is cosmological in origin. Whether this signature arises from previously 
proposed sources of non-Gaussianity, such as cosmic strings, large spots, 
matter-antimatter domain interfaces, etc., remains to be determined.

Recall that the COBE-DMR data tolerate almost all popular models of
primordial density perturbations in terms of the  second order statistics.
Generally, the data are only able to discriminate among the power spectra of 
these models with less than 2-$\sigma$ confidence levels \cite{ji94}. The 
scale-scale correlation detected in the 4-year COBE-DMR data either gives a 
rather high confidence of ruling out the CDM model or evidence for the 
existence of unknown local foreground contamination on angular scales as 
large as $\simeq$ 10 -- 20 degrees. Obviously, if either of these implications
are correct, two important conclusions can be inferred: 1.) the inflation
plus cold dark matter model with standard cosmological parameters appears 
to be ruled out at the $>99\%$ confidence; 2.) the COBE-DMR temperature maps 
are contaminated on large angular scales at levels larger than previously 
thought. Whether COBE determined cosmological parameters, such as the 
quadrupole of temperature fluctuations, may also be contaminated remains to 
be seen. 

We are very grateful to Al Kogut for providing the systematic error and 408 MHz
maps used in the analysis.

\begin{table}
\caption{Measured $C_j^{2,2}$ coefficients.}
\begin{tabular}{clcl|ccl}
\multicolumn{3}{r}{ FACE 0}&& \multicolumn{3}{c}{ FACE 5} \\ \hline 
$j$ & DMR & CDM+DSMB & 95\% Confidence Levels& DMR&CDM+DSMB& 95\% Confidence Levels\\
 \tableline
 1  & 2.091 &  1.004 & (0.376 -- 1.587) &0.730&1.008& (0.492 -- 1.761) \\ 
 2  & 0.984 &  1.035 & (0.601 -- 1.513) & 1.300&1.022&(0.688 -- 1.572)\\ 
 3  & 1.041 &  1.032 & (0.791 -- 1.294) & 1.172& 1.026&(0.832 -- 1.358) \\
\end{tabular}
\label{table1}
\end{table}

\newpage

$ \; $ \\

\begin{center}
\begin{figure}
\epsfxsize=12.0cm
\begin{turn}{-90}
\centerline{\epsfbox{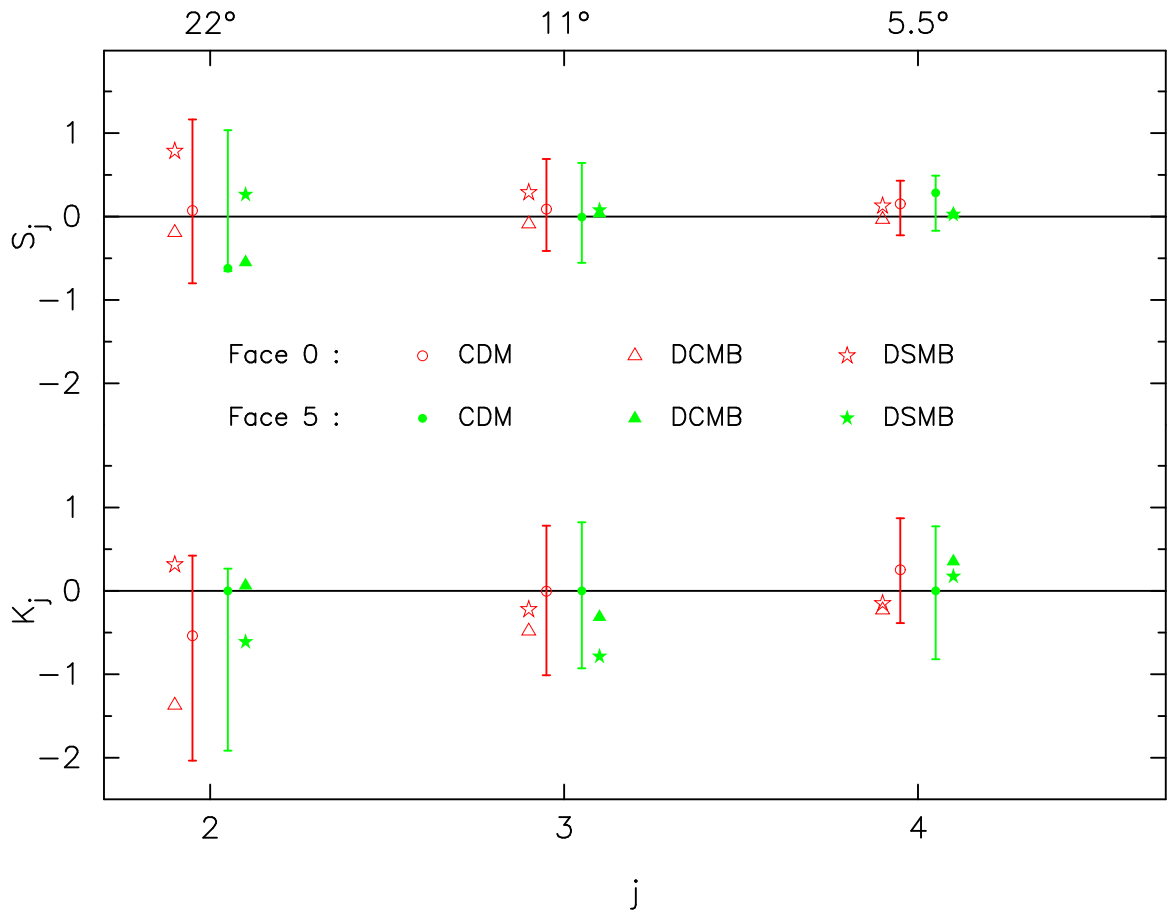}}
\end{turn}
\vskip 7mm
\caption{$S_j$ (top) and $K_j$ (bottom) for faces 0 and 5 of the 
DMR data and the CDM simulations. } 
\label{fig 1}
\end{figure}
\end{center}

\newpage

$ \; $ \\
\vskip 3.0cm

\begin{center}
\begin{figure}
\epsfxsize=12.0cm
\centerline{\epsfbox{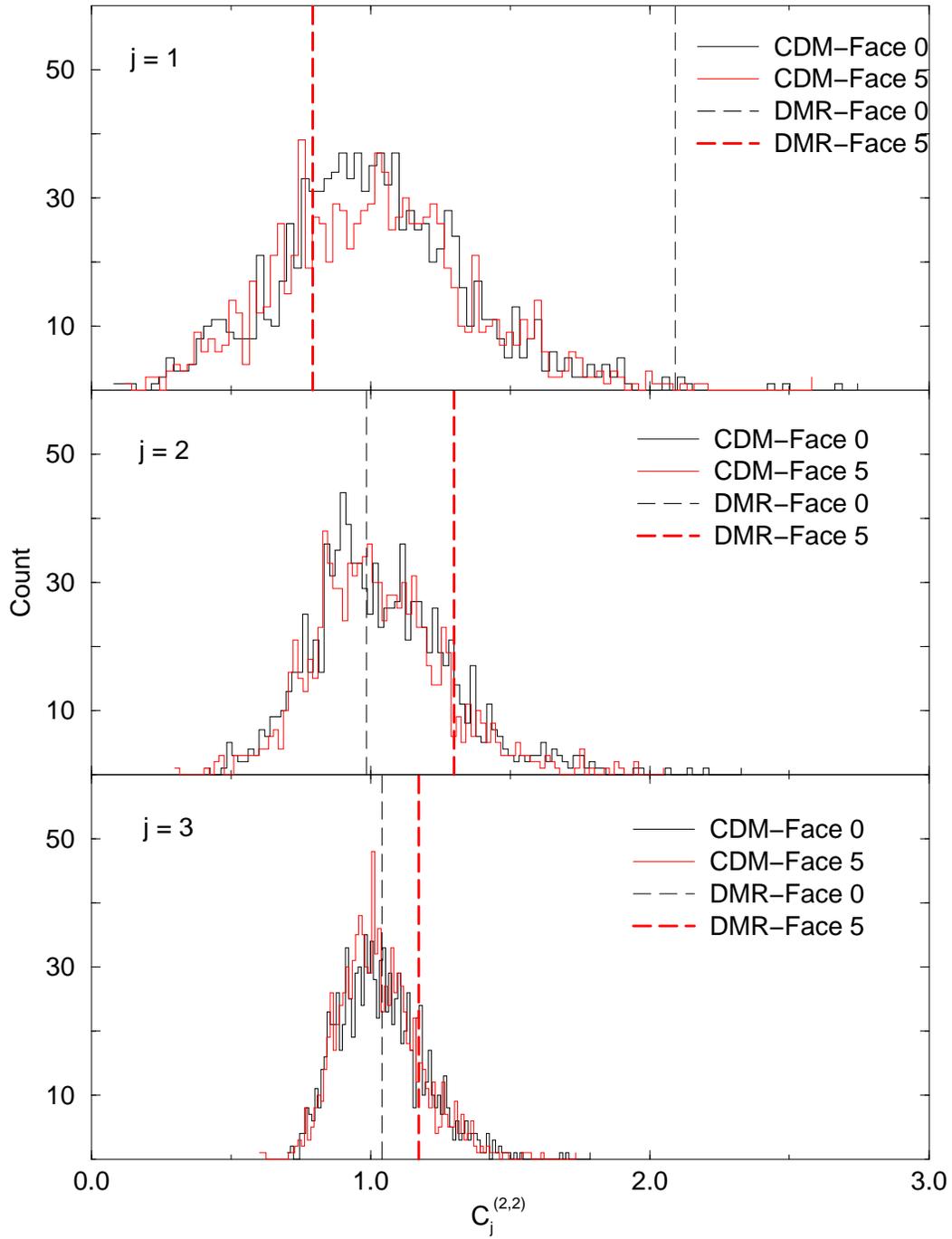}}
\vskip 7mm
\caption{$C_j^{(2,2)}$ for faces 0 and 5 of the DMR data and the CDM
simulations.} 
\label{fig 3}
\end{figure}
\end{center}

\end{document}